# THE MISSING MEASUREMENTS OF THE GRAVITATIONAL CONSTANT

Maurizio Michelini

ENEA – Casaccia Research Centre – Rome , Italy
maurizio.michelini@casaccia.enea.it *

**Abstract –** In 1998 - two centuries after Cavendish - a conference on theory and experiment of the G measurement pointed out the progress made in various experimental methods and discussed the effects on the accuracy of G. In spite of several measurements with torsion balance in "vacuum" to the aim of reducing some disturbances (air thermal convection, aerodynamic resistance upon the moving pendulum, etc.), no mention was made about a possible pressure effect in calm air. In 2000 J.Luo and Z.K.Hu firstly denounced the presence of some unknown systematic problem on G measurement. In the present work a new systematic error is analysed which arises from the non-zero balance of the overall momentum discharged by the air molecules upon the test mass within the vacuum chamber. This effect is normally negligible, but when the pressure is so low that the molecule mean free path is comparable to the thickness of the air meatus surrounding the test mass , the drawing force may become greater than the gravitational force. Considering the usual size of the meatus, the molecular effect becomes maximum when the pressure drops to about $10^{-5}$ bar. Before Heyl's measurement at 1 millibar (1927), the experiments were made at higher pressures. Conversely those made with recent vacuum techniques show pressures down to $10^{-10}$ bar (Gundlach & Merkowitz, 2000) and $10^{-11}$ bar (Gershteyn, 2002). In these experiments the effect of the "vacuum" pressure appears very little. As a matter of fact, we were not able to find in the literature some measurements made at vacuum pressures between the millibar and the nanobar. Why ? This lack appears embarrassing in absence of an adequate physical explanation.

* new adress : 317michyz28@libero.it

## 1. Introduction

Everyone knows the simple experience of two flat microscopy glasses which cannot be separated from each other when their surfaces touch.
Obviously this effect is due to the external pressure of the air molecules which are unable to penetrate between the corrugations of the polished surfaces, so within the small meatus there is a considerable air depression. The mean free path of the air molecules at normal pressure is about $10^{-7}$ metres, that is of the same order of magnitude of the polished surface corrugations. In general, the molecules are not able to *freely* penetrate within a meatus whose thickness is reduced to about 1 *mean free path* .
 When we consider the meatus facing the gravitational test mass of a torsion balance placed in a vacuum chamber, the air depression within the meatus originates a little drawing force which adds to the gravitational force. The disturbing force is often negligible, but when the pressure within the vacuum chamber is reduced beyond a millibar (to avoid other disturbances due to air convection) the meatus optical thickness reduces, so the above condition about 1 mean free path may occur.
 As a consequence, it appears necessary to investigate this phenomenon to obtain a quantitative prediction of this systematic drawing force arising in the G measurements.
 The investigation shows that the total momentum discharged by the non-isotropic flux of gas particles upon two masses separated by a small meatus, originates a drawing force.
 In paragraph 7 we examine the case of a uniform flux of small quantum waves which originate between two masses a drawing force reproducing the same characteristics of the Newton's gravitational force . This new paradigm puts new light on the transmission of the gravitational interaction, a problem existing both in Newon's theory as well as in general relativity.

## 2. Measurements of the gravitational constant G

   The torsion balance apparatus was first used by Cavendish in 1798 in a simple form which permitted him to reach an unexpected accuracy. In the following two centuries the torsion balance was used by several experimenters (Boys, Eotvos, Heyl, etc.) who substantially improved the technique, but the level of accuracy did not show a dramatic enhancement.
   Several methods were devised in the XX century to measure G.
  In a Conference organised in London in 1998 – two centuries after Cavendish - by C.C.Speake and T.J.Quinn [1] a variety of papers described the methods of measurement and their potential accuracy related to the disturbances and systematic errors. Many experiments were described. For instance: a torsion balance where the gravitational torque is balanced by an electrostatic torque produced by an electrometer; a torsion-strip balance where the fibre is substituted by a strip; a dynamic method based on a rotating torsion pendulum with angular acceleration feedback; a *free fall* method where the determination of G depends on changes in acceleration of the falling object, etc.
   Notwithstanding the technological improvement, up to now the gravitational constant is the less accurately known among the most important constants in physics.
   The uncertainty has been recognised to depend on various experimental factors.

To eliminate the air thermal convection on the test spheres, in 1897 K.F. Braun made a torsion balance measurement after extracting the air from the ampoule. The level of vacuum obtained with the techniques of the time is not known.

   In 1927 Heyl [2] made a benchmark measurement with a heavy torsion balance to the aim of establishing a firm value of G. Since nobody suspected a possible disturbance due to the calm air, the pressure $p_o$ was reduced to 1-2 millibar, a level sufficient to eliminate the air convection currents.
   The molecule mean free path at that pressure is about $10^{-4}$ metres, a quantity much smaller than the thickness of the meatus. From our investigation it appears that the air pressure effect does not alter the accuracy of the classical G measurements performed at pressures higher than some millibars.
    In 1987 G.T. Gillies published an Index of measurements [3] containing over 200 experiments, "but many of them have not been reported in what would now be considered to be the open literature".

   A status of the recent G measurement was published in 2000 by J.Luo and Z.K.Hu [4] in which the presence of some unknown systematic problem was first denounced : *"the last nine measurements have produced values of G that differ wildly from each other"*.
   In 2003 the situation was analysed by R.Kritzer [5] who concluded that *"the large spread in results compared to small error estimates, indicates that there are large systematic errors in various results"*.

   Among the last experiments, some of them used new sophisticated methods with technologies coupled to *very low* pressures within the test chamber.
   J.H. Gundlach and S.M.Merkowitz [6] made a measurement where a flat pendulum is suspended by a torsion fiber without torque since the accelerated rotation of the attracting masses equals the gravitational acceleration of the pendulum. To minimise the air dynamic resistance, the pressure was lowered to $10^{-7}$ Torr ($p_o \approx 10^{-10}$ bar).
   At this pressure the usual molecular mean free path $l = m/\sigma\,\delta_o$ (valid within a large homogeneous medium) is of the order of 1000 metres. Hence within the vacuum chamber the lack of flux homogeneity is everywhere present and becomes maximum within a small meatus.

Another accurate measurement was performed in 2002 by M.L.Gershteyn et al. [7] with the *time of swing* method, where the pendulum feels a unique drawing mass fixed at different distances from the test mass. The change of the oscillation period determines G. To minimise the air disturbance, the pressure in the vacuum chamber was lowered to $10^{-6}$ Pascal (i.e. $p_o = 10^{-11}$ bar). Also in this experiment the reason for such a dramatic lowering of the air pressure is not discussed. Notice that the air dynamic resistance to the motion of the test mass is of the order of $10^{-25}$ newton, a quantity totally negligible compared to the measured gravitational force, estimated between $10^{-11} \div 10^{-10}$ newton. The authors revealed the presence of a variation of G with the orientation (regard to the fixed stars) amounting to 0.054%. Incidentally, the anisotropy of G agrees with the gravitational-inertial theory discussed in paragr. 7.

In 2004 a new torsion balance configuration with four attracting spheres located within the vacuum chamber ($p_o = 1.5 \times 10^{-10}$ bar) was described by Z.K.Hu and J.Luo [8]. The four masses are aligned and each test mass oscillates within the large meatus (about 4 cm) between a pair of attracting masses. Each test mass determines with the adjacent spheres a small meatus (estimated about 4 mm) and a large meatus (about 16 mm). During the experiment the authors found the presence of an abnormal period of the torsion pendulum, which resulted independent of the material wire, test mass, torsion beam and could not be explained with external magnetic or electric fields. Adopting a magnetic damper system, the abnormal mode was suppressed, but the variance of the fundamental period of the pendulum introduced an uncertainty as large as 1400 ppm, testifying the presence of a systematic disturbance in determining G.

We applied to this problem the analysis carried out in this paper (paragr. 4-5). From the air density of the vacuum chamber, we calculate the optical thickness of the small meatus and the related air depression (eq.6), which substituted in eq(10) gives upon the test mass a disturbing force rising up to $F(p_o) \approx 10^{-14}$ newton, equivalent to about $10^{-4}$ times the gravitational force, which alters the pendulum period. This fact agrees with the author conclusions [8] that the torsion balance configuration would have an inherent accuracy of about 10 ppm in determining G, but the uncertainty in the fundamental period reduces this accuracy to 1400 ppm.

The presence of an abnormal disturbance was previously described (1998) by Z.K.Hu, J.Luo, X.H. Fu et al. [9] in dealing with the time-of-swing method. They found the presence of "*important nonlinear effects in the motion of the pendulum itself, independent of any defect in the detector, caused by the finite amplitude of the swing*". Their configuration consisted in a torsion balance with heavy masses external to the vacuum chamber, where the pressure was lowered to $p_o = 2 \times 10^{-10}$ bar. The test mass, diameter about 19 mm, was suspended within a stainless vacuum tube placed between two heavy masses distant 60 mm apart. Since the test mass oscillates up to 8 mm from the centre of the vacuum tube, the optical thickness of the small meatus can be deduced. The smaller this thickness, the greater the disturbing force $F(p_o)$. Repeating the analysis carried out for the preceding experiment, we found a force $F(p_o)$ which represents a little fraction of the gravitational force, due to the heavy attractor masses.

Comparing with many measurements done in the era of high vacuum technology [10,11,12,13,14,15,16,17,18] we observed that the experimenters avoided to use in the vacuum chambers pressures between the millibar and the nanobar. The reason for this avoidance does not appear to have been discussed.

## 3. The scattering of gas particles upon smooth surfaces

The scattering of molecules hitting a smooth surface does not generally follow the optical reflection because they interact mainly upon single atoms/molecules of the lattice. Like the random scattering arising when two free particles come in collision, the nearly right-angled molecules hitting a surface can be scattered in all directions. Conversely, the molecules hitting the surface from a nearly parallel direction interact softly with the field of the atomic lattice. In fact the nearly parallel molecules, whose momentum $\mathbf{q} = m\mathbf{v}$ makes an angle $\alpha \approx \pi/2$ with the vertical axis, receive from the field a small vertical momentum $\Delta q \approx 2m v \cos\alpha$ which redirects the molecules along a nearly optical reflection. Like the UV photons (whose momentum $h\nu/c$ is comparable to the momentum of air molecules at normal temperature) even the air molecules show a limiting angle of optical reflection on a polished surface.

To resume: after scattering on a flat surface the quasi-parallel molecules remain as such, whereas a fraction of the nearly orthogonal molecules become quasi-parallel.
As a consequence an isotropic flux of molecules hitting a smooth surface, after scattering becomes a non-isotropic flux $\psi_w(\alpha)$. This condition may be described by the relationship

$$(1) \qquad \psi_w(\alpha) \cong \phi_o (1 - \Delta_1 \cos\alpha + \Delta_2 \sin\alpha)$$

where $\phi_o$ is the incident flux and the parameters $\Delta_1, \Delta_2$ satisfy the condition

$$\int_o^{\pi/2} \sin\alpha \, \psi_w(\alpha) \, d\alpha = \phi_o .$$

Moreover we assume (similarly with the scattering of the photons) that about 10% of the nearly right-angle molecules become quasi-parallel after scattering on the wall. Applying these two conditions one obtains the figures $\Delta_1 \cong 0.146$, $\Delta_2 = 2\Delta_1/\pi \cong 0.0928$. Different numerical assumptions do not dramatically change these coefficients.

This physical condition is important to understand the phenomenon of the molecular flux depression within the meatus between the moving mass and the adjacent vacuum chamber. This phenomenon happens at low air pressures. When the pressure is higher than a millibar, within the meatus more than 99.99% of the molecules hitting the moving mass come from scattering with other molecules, whereas less than $10^{-4}$ molecules come directly from the scattering on the chamber side. To feel a little flux depression in the meatus it is necessary that the molecules coming from surface-scattering be about a half of the total. Within an air meatus of thickness $s$ this happens when the optical thickness $\Sigma s = s\sigma \delta_o/m \approx 10^7 s \, \delta_o$ equals about 1 mean free path, i.e. when the air density equals $\delta_o \approx 10^{-7}/s$.
For usual torsion balances the critical vacuum pressure is around $p_o \approx (1 \div 3) \times 10^{-5}$ bar.

## 4. The molecule momentum discharged on the gravitational mass

The old G measurements adopted a torsion balance at atmospheric pressure, that is without the vacuum chamber, so the pressure effect took place in the meatus between the test mass and the attracting sphere. This happens also in G measurements in vacuum when the heavy masses are comprised within the chamber, but in general the G measurements in vacuum are made with the heavy masses outside the chamber. In this case we define "meatus" the volume of air comprised between the test mass and the adjacent wall of the vacuum chamber (fig.1).
To the aim of reducing the disturbances due to the air convection, in 1897 the torsion balance was placed by Braun within a vacuum chamber where the pressure was reduced to the level allowed by the vacuum technologies of the time. At a pressure of the order of some millibar we found that the molecular flux upon the moving mass results highly uniform, so the sum of every momentum discharged by molecules on the sphere is null for any practical purpose.

However, when the pressure in the chamber is strongly reduced, the molecular flux shows a little depression in the meatus. The flux in the circular meatus may be expressed along the radial direction $x$ as follows

$$\phi(x) \cong \phi_m(1 + k x^2) \qquad (2)$$

where $\phi_m$ is the figure the flux takes on the axis. The flux on the meatus boundary ($x = L$) is the unperturbed flux $\phi_o$, so $\phi_m(1 + k L^2) = \phi_o$, which shows that $k$ is linked to the flux depression

$$k = (\phi_o/\phi_m - 1)/L^2 \qquad (2a)$$

where $L \cong R \cos\beta$ is the radius of the area of the moving mass experiencing the flux depression. The angle $\beta$, defined by $\sin\beta = R/R+s$ (where $R$ is the radius of the moving mass, $s$ is the minimum thickness of the meatus), plays a fundamental role in the phenomenon since it describes the "shadow" of the moving mass on the chamber wall.

Choosing spherical co-ordinates with the same axis of the meatus and origin (fig.1) in the point B, the following equation gives us the angular flux $\psi_B(\alpha)$ of incident molecules integrating along the thickness $s(\alpha)$ the scattered molecules, to which one must add the uncollided molecules due to the flux $\psi_S(\alpha)$ scattered on the surface of the moving mass

$$\psi_B(\alpha) = \int_o^{s(\alpha)} \Sigma \phi(r) \exp(-\underline{\Sigma} r)\, dr + \psi_S(\alpha)\, \exp(-\underline{\Sigma} s(\alpha)) \qquad (3)$$

where $\Sigma(r) = \sigma\, \delta(r)/m$ is the air macroscopic cross section; $\Sigma \phi(r)$ is the density of the isotropically scattered molecules; $\underline{\Sigma} s(\alpha)$ is the optical thickness along $\alpha$ of the meatus. This angular flux holds for $\alpha \leq \beta$.

This presentation of the problem has only an instructive character, because the fluxes $\phi(r)$ and $\psi_S(\alpha)$ are unknown.

Fig. 1 – Schematic drawing of a torsion balance comprised within a vacuum chamber.

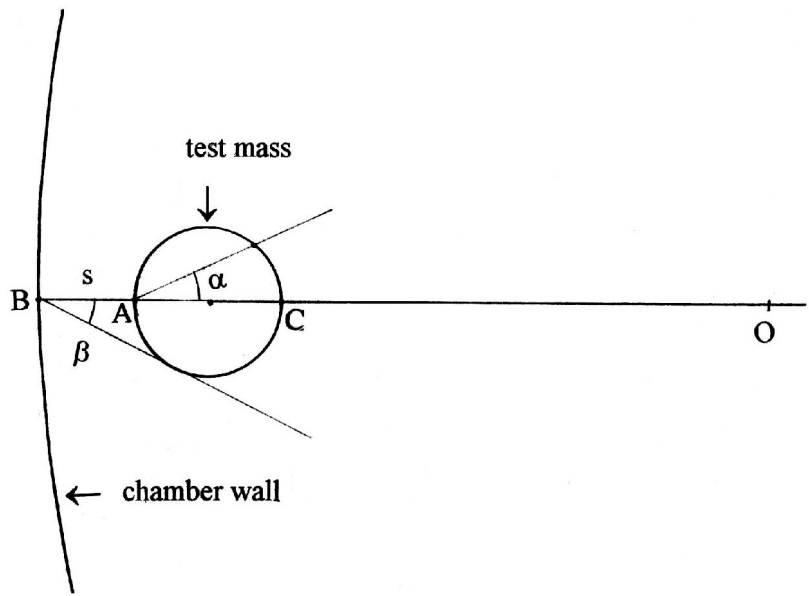

*The principle of superposition of the effects*

To solve the problem of calculating the molecular flux within the meatus we adopt the principle of superposition of the effects. Consider the test sphere surrounded by the air comprised in the vacuum chamber at pressure $p_o$. To obtain the disturbing force F($p_o$) on the test mass related to the air depression in the meatus we must calculate the flux in the point A of the sphere and in the point C diametrically opposite (fig.1). Let's now *remove* the sphere and substitute an equal volume of air at pressure $p_o$. In this case the chamber results filled with the uniform flux $\phi_o$. Now we consider the spherical co-ordinates with origin in the point A (fig.1). The virtual flux $\psi'_r$ on the right of the point A comes from scattering on the air molecules comprised in the whole chamber (whose size equals $n \gg 1$ m.f.p.)

$$(3) \qquad \psi'_r = \int_o^n \Sigma \phi_o \exp(-\Sigma r)\, dr \; .$$

From the formal standpoint, eq(3) may be rewritten as follows

$$(3a) \qquad \psi'_r = \int_o^{t(\alpha)} \Sigma \phi_o \exp(-\Sigma r)\, dr + \phi_o \exp(-\Sigma t(\alpha))$$

where $t(\alpha) = 2R \cos\alpha$ is the distance between the points A and P (fig.1) placed on the virtual surface of the removed mass. Let's notice that the first term in eq(3a) represents the flux due to the air scattering source occupying the sphere volume.

When we cancel this source term (for instance reintroducing the test mass), eq(3a) gives the flux

$$(3b) \qquad \psi_{Ar}(\alpha) = \phi_o \exp(-2\Sigma R \cos\alpha) \; .$$

On the left of the point A the flux comes from scattering in the region of the air meatus

$$(3c) \qquad \psi_{Al}(\alpha) = \int_o^{z(\alpha)} \Sigma \phi_o \exp(-\Sigma r)\, dr + \psi_w(\alpha) \exp(-\Sigma z(\alpha)) =$$

$$= \phi_o\, [1 - \exp(-\Sigma z(\alpha))] + \psi_w(\alpha) \exp(-\Sigma z(\alpha))$$

where $z(\alpha)$ is the wall distance and $\psi_w(\alpha)$ is the flux scattered on the chamber wall, as defined by eq(1). Since in general the radius of the chamber is much larger than $R$, one may assume the distance $z(\alpha) \cong s/\cos\alpha$. Subtracting the flux $\psi_{Ar}(\alpha)$ from the flux $\psi_{Al}(\alpha)$, one gets the actual flux hitting on the point A of the sphere

$$(4) \qquad \psi_A(\alpha) \cong \phi_o\, [1 - \exp(-2\Sigma R \cos\alpha)] - [\phi_o - \psi_w(\alpha)] \exp(-\Sigma s/\cos\alpha).$$

Now we calculate with the same procedure the flux hitting on the point C

$$(4a) \qquad \psi_C(\alpha) \cong \phi_o\, [1 - \exp(-2\Sigma R \cos\alpha)] - [\phi_o - \psi_w(\alpha)]\, \exp(-\Sigma(s+2R)/\cos\alpha)].$$

The disturbing force acting on the moving mass is linked to the different pressures due to the molecule momentum discharged near the points A and C. The pressure is related to the molecular flux, which shows the following difference across the test mass diameter

$$\Delta\phi_o/\phi_o = (\phi_C - \phi_A)/\phi_o = \int_o^{\pi/2} \sin\alpha\, [\psi_C(\alpha) - \psi_A(\alpha)]d\alpha \; .$$

Substituting and putting $w(\alpha) = \psi_w(\alpha)/\phi_o$, one gets the flux difference

$$(5) \qquad \Delta\phi_o/\phi_o = \int_o^{\pi/2} \sin\alpha\, [1 - w(\alpha)]\, [\exp(-\Sigma s/\cos\alpha) - \exp(-\Sigma(s+2R)/\cos\alpha)]d\alpha$$

which confirms that the meatus depression depends on the anisotropy of the flux $\psi_w(\alpha)$ scattered on the wall. Through eq(1) we also have

$$w(\alpha) = 1 - \Delta_1 \cos\alpha + \Delta_2 \sin\alpha$$

which, substituting in eq(5) gives the central air depression

(6)     $\Delta p_o/p_o = \Delta \phi_o/\phi_o = [\Delta_1 \Gamma(\Sigma s, \Sigma R) - \Delta_2 \Omega(\Sigma s, \Sigma R)]$

where the functions

(6a)     $\Gamma(\Sigma s, \Sigma R) = \int_0^{\pi/2} \sin\alpha \cos\alpha \, [\exp(-\Sigma s/\cos\alpha) - \exp(-\Sigma(s+2R)/\cos\alpha)] d\alpha$

(6b)     $\Omega(\Sigma s, \Sigma R) = \int_0^{\pi/2} \sin^2\alpha \, [\exp(-\Sigma s/\cos\alpha) - \exp(-\Sigma(s+2R)/\cos\alpha)] \, d\alpha$

depend on the meatus geometry and on the air density $\delta_o = \Sigma m/\sigma$ in the vacuum chamber. These functions do not appear to have been tabulated. Approximate fitting functions have been used for calculations, whose accuracy does not appear completely satisfying.
To give a quantitative idea, the relative depression $\Delta p_o/p_o$ had been calculated assuming figures related to the usual size of a torsion balance. For instance:

- Physical parameters : moving mass radius $R = 5$ mm, meatus minimum thickness $s = 4$ mm, chamber pressure $p_o = 1$ millibar, air density $\cong 10^{-3}$ kg/m$^3$, macroscopic scattering cross section $\Sigma \cong 10^4$ m$^{-1}$, $\Sigma s \cong 40$.

Substituting in eq(6) one obtains $\Delta p_o/p_o \approx 4.5 \times 10^{-19}$ which shows the high uniformity of the molecular flux within the meatus at a pressure of 1 millibar.

However the chamber pressure $p_o = 10^{-5}$ bar corresponds a sensible depression $\Delta p_o/p_o \approx 3.37 \times 10^{-3}$ which largely alters the measured gravitational force.

## 5. The calculation of the drawing force

The drawing force on the moving mass due to the small lowering of the air pressure $p(r) = mv\phi(r)$ within the meatus, is defined by

(9)     $F = \int_0^L 2\pi r \, [p_o - p(r)] \, dr$

where $L = R \cos\beta$ is the radius of the circular section of the meatus where $p(L) = p_o$.

Substituting the flux distribution given by eq(2) one gets

$$p(r) = mv\phi_m (1 + k r^2)$$

and the corresponding depression within the meatus

$$p_o - p(r) = p_o [1 - (\phi_m/\phi_o)(1 + k r^2)].$$

Substituting the expression of $k$ by eq(2a) one obtains

$$p_o - p(r) = p_o (1 - \phi_m/\phi_o)(1 - r^2/L^2)$$

which, substituted in eq(9), gives us the force

$$F = (\pi/2) \, p_o L^2 \, \Delta\phi_o/\phi_o = (\pi/2) \, p_o L^2 \, \Delta p_o/p_o$$

where $L = R \cos\beta$ and the relative depression is given by eq(6).
Substituting we obtain the drawing force due to the air effect when the pressure in the vacuum chamber is $p_o$

(10)     $F = (\pi/2) \, p_o (R \cos\beta)^2 \, [\Delta_1 \Gamma(\Sigma s, \Sigma R) - \Delta_2 \Omega(\Sigma s, \Sigma R)]$.

To give an idea of this force we calculated in Table 1 some figures of $F(p_o)$ for the same torsion balance apparatus previously examined.

## 6. Discussion of the numerical results

From the results reported on Table 1 one can notice that in the assumed torsion balance apparatus (with light moving mass, R = 5 mm) the drawing force $F(p_o)$ takes a maximum at a pressure $p_o \approx 10^{-5}$ bar which makes the optical thickness of the meatus about equal to 1.

This maximum $F_M$ is estimated to rise well over the figure of the measured gravitational force $F_{gr}$ (in general between $10^{-10} \div 10^{-11}$ newton). However in the Gershteyn's light torsion balance the measurement was made at a pressure $p_o = 10^{-11}$ bar, so the disturbing force $F(p_o)$ results about $10^{-7}$ times the measured gravitational force.

In the Heyl's heavy balance experiment the measured $F_{gr}$ was of the order of $10^{-9}$ newton and the maximum $F_M$ rises over $F_{gr}$. However the disturbing force $F(p_o)$ at a pressure $p_o = 1$ millibar results about $10^{13}$ times lower than $F_{gr}$.

These few data explain *ad abundantiam* why the region of the intermediate pressures between the millibar and the nanobar was avoided by the experimenters.

Obviously, what is of interest in the measurements of G is the possible systematic error due to $F(p_o)$. For instance in the case of the advanced measurement of Gundlach & Merkowitz [3] within a vacuum chamber at $p_o \approx 10^{-10}$ bar, the force $F(p_o)$ is estimated to be less than $10^{-5}$ times the measured gravitational force.

Table 1 is linked to a fixed size of the meatus and test mass. In practice the meatus thickness may rise to $s \approx 10$ mm. As a consequence the force $F(p_o)$ results $2 \div 3$ times lower than reported in the Table (at equal figures of $\Sigma s$) since the maximum $F_M$ is displaced toward pressures $2 \div 3$ times lower than $p_o$.

Table 1 – Calculation of the drawing force due to the air m0olecules within the vacuum chamber of a gravitational torsion balance. The assumed geometrical characteristics are : meatus thickness $s = $ 4mm, moving mass radius $R = 5$mm.

| *Vacuum pressure* $p_o$ (pascal) | *Air density* $\delta_o$ (kg/m$^3$) | *Meatus optical thickness* $\Sigma s$ (m.f.p.) | *Molecular flux depression in the meatus* $\Delta\phi_o/\phi_o$ | *Disturbing drawing force* $F(p_o)$ (newton) |
|---|---|---|---|---|
| 100 | $10^{-3}$ | 40 | $1.4 \times 10^{-20}$ | $1.0 \times 10^{-23}$ |
| 10 | $10^{-4}$ | 4 | $2.86 \times 10^{-4}$ | $7.2 \times 10^{-8}$ |
| 1 | $10^{-5}$ | 0.4 | $3.37 \times 10^{-3}$ | $8.4 \times 10^{-8}$ |
| 0.1 | $10^{-6}$ | $4 \times 10^{-2}$ | $6.74 \times 10^{-3}$ | $1.7 \times 10^{-8}$ |
| $10^{-2}$ | $10^{-7}$ | $4 \times 10^{-3}$ | $1.8 \times 10^{-3}$ | $4.5 \times 10^{-10}$ |
| $10^{-3}$ | $10^{-8}$ | $4 \times 10^{-4}$ | $\approx 4.4 \times 10^{-4}$ | $\approx 1.1 \times 10^{-11}$ |
| $10^{-4}$ | $10^{-9}$ | $4 \times 10^{-5}$ | $\approx 1.1 \times 10^{-4}$ | $\approx 2.8 \times 10^{-13}$ |
| $10^{-5}$ | $10^{-10}$ | $4 \times 10^{-6}$ | $\approx 2.8 \times 10^{-5}$ | $\approx 7 \times 10^{-15}$ |
| $10^{-6}$ | $10^{-11}$ | $4 \times 10^{-7}$ | $\approx 8 \times 10^{-6}$ | $\approx 2 \times 10^{-16}$ |

## 7. The interaction of a quantum wave flux with matter

The Newton's action at a distance, as well as the experimental detection of the relativistic gravitational waves (which encounters great difficulties), are perhaps the most puzzling problem of the present-day physics. According to G.T.Gillies [3], we think that general relativity constitutes "the modernized presentation" of the classical gravitational interaction, which did not receive substantial advancement after Newton. Some authors firmly deny [19] that general relativity predicts gravitational waves produced by the motion of bodies.
The problem of the transmission of any physical interaction is linked to the following great paradigms:

1) the well experienced field theory requires that any force between two charges or two masses be transmitted by some kind of waves, which are specific to the interaction,

2) the Newton's "gravitational mass" is an old paradigm which was accepted without an adequate criticism in general relativity theory. It works well in the ordinary calculations, but does not explain (for instance) the problem of the galactic supermassive obscure bodies [20] and generates difficulties in observing the waves imagined to transmit the interaction between the "gravitational masses".

Is there anything which restricts the solution of this problem to the experienced paradigms? We think not. At any time the research finds new paradigms which interpret the physical evidences better than the preceding ones.

The gravitational force between bodies is a physical evidence. How can a mass feel the force due to another distant mass? From the epistemology point of view we have to choose between

- accepting a force between distant "gravitational masses" transmitted by the deformation of the *void* space; however the void space does not constitute a *physical* reality interacting with other *physical* realities,

- accepting a gravitational force quantitatively linked to each mass, but generated from the action of a surrounding *physical* agent, such as a flux of particles or quantum waves.

There is no doubt that the second choice is less embarrassing because we know interactions where the force between two bodies is generated by an external surrounding flux (particles, waves, etc.) modified by the *mutually shielding* bodies.

In this paper, dedicated to the analysis of G measurement, the generation of a drawing force on the torsion balance has been shown quantitatively to depend on the air density in the vacuum chamber. The non-isotropy of the air molecular flux hitting the test mass gives rise to an *unknown* disturbing force, which in certain low pressure conditions becomes greater than the measured gravitational force.

This "strange" drawing force was not recognised during the past two centuries because it is contrary to common belief. The G experimenters who encountered this disturbance avoided discussing the problem, preferring to adopt conditions of measurement where it is negligible. The problem was not solved, but removed.

*Reflections about the gravitational-inertial force*

Part of the physicists prefer the "economic" old concept of gravitational-mass instead of joining the complex gravitational force based on the interaction of matter with an isotropic flux of very small waves (cosmic quanta) filling the space. This paradigm requires a gravitational constant depending on the orientation respect to the fixed stars [21], in accord with the G measurement made in 2002 by M.Gershteyn et al.[6]. These waves have a little energy $E_o$ and a wavelength $\lambda_o$ which results to be equal to the Planck's length, while the relevant quantum constant $h_o = E_o\lambda_o/c$ is much smaller than the usual Planck's constant $h$. The flux of the cosmic quanta was demonstrated to be the physical reality responsible of the relativistic inertial forces [22]. The very high accuracy of the related equations (which induced to believe that the inertial forces depend on the space-continuum) is due to the very high number of *contemporaneous* collisions upon a particle. As a consequence of the very high flux, the wavelength of the cosmic quanta, which do not collide among them, results very small.

The inertial mass of the elementary particles is proportional to the cross section $\sigma$ they show in the Compton collisions with the cosmic quanta [22]. As a consequence, $\sigma/m = A_o$ is a constant for any particle. Differently from classical and relativistic physics, which "explain" the inertial forces through the motion of particles through the void space, the laws of motion depend on the collisions between the cosmic quanta and matter.

The cosmic quanta easily cross the celestial bodies (excepting the very dense neutron stars) leaving on each mass a little momentum proportional to the shield due to an other mass. This momentum is responsible for a *drawing* force linked exactly to the inverse square distance, because a mass of ($M/m$) nucleons with inertial cross section $\sigma$ makes (at a distance $r$) a shield towards the cosmic flux related to the solid angle

$$\gamma(r) = M\sigma/m\, 2\pi r^2 = A_o M /2\pi r^2$$

which originates the Newton's gravitational force linked to the constant

$$G = K_o \phi_o E_o A_o^2/4\pi c$$

depending on the cosmic quanta characteristics [22].

The principle of equivalence states that inertial and gravitational mass are equal, as experimentally verified with an exceedingly high degree of accuracy. Is there a link between the above gravitation theory and this principle ? The answer is that the theory verifies entirely the strong principle of equivalence, since *both* inertia and gravitation *descend from the same physical phenomenon,* i.e. the collision process between the elementary particles and the cosmic quanta flux. The elementary particles of each mass receive, colliding with the flux, the inertial forces, whereas the flux modification due to the *mutual* shielding between masses generates a drawing force which is known as the gravitational force.

Someone may ask: Why leave the paradigm of "gravitational-mass" when the practical effects of the two paradigms appear to be similar ?
There are many reasons to do this. The most basic one is that the old "gravitational-mass" undergoes the embarrassing *unlimited* collapse, which produces the *infinite* compression of matter. This inconsistency rules out of physics the *unlimited* collapse.
As a consequence, rational science can do nothing else but to reject even the *gravitational-mass* paradigm..